\documentstyle[12pt]{article}

\font\tenrm=cmr10
 at 10truept

\def\ga{\gamma}

\def\si{\sigma}

\def\ps{\psi}
\def\om{\omega}

\def\De{\Delta}

\def\fr#1#2{{{#1} \over {#2}}}

\def\half{{\textstyle{1\over 2}}}
\def\frac#1#2{{\textstyle{{#1}\over {#2}}}}

\def\lsim{\mathrel{\rlap{\lower4pt\hbox{\hskip1pt$\sim$}}
    \raise1pt\hbox{$<$}}}
\def\gsim{\mathrel{\rlap{\lower4pt\hbox{\hskip1pt$\sim$}}
    \raise1pt\hbox{$>$}}}
\def\sqr#1#2{{\vcenter{\vbox{\hrule height.#2pt
         \hbox{\vrule width.#2pt height#1pt \kern#1pt
         \vrule width.#2pt}
         \hrule height.#2pt}}}}

\newcommand{\beq}{\begin{equation}}
\newcommand{\eeq}{\end{equation}}
\newcommand{\bea}{\begin{eqnarray}}
\newcommand{\eea}{\end{eqnarray}}
\newcommand{\rf}[1]{(\ref{#1})}

\renewenvironment{thebibliography}[1]
 { \rm
   \begin{list}{\arabic{enumi}.}
    {\usecounter{enumi} \setlength{\parsep}{0pt}
     \setlength{\itemsep}{3pt} \settowidth{\labelwidth}{#1.}
     \sloppy
    }}{\end{list}}

\begin{document}

\baselineskip=16pt
\begin{flushright}
{COLBY 98-02\\}
{IUHET 385\\}
{February 1998\\}
\end{flushright}
\vglue 1.0 truein 

\begin{flushleft}
{\bf TESTS OF CPT AND LORENTZ SYMMETRY\\
 IN PENNING-TRAP EXPERIMENTS
\footnote{\tenrm
Presented by R.B. at Orbis Scientiae: Physics of Mass, 
Miami, Florida, December 1997}
\\}
\end{flushleft}

\vglue 0.8cm
\begin{flushleft}
{\hskip 1 truein
Robert Bluhm$^a$, V. Alan Kosteleck\'y$^b$, and Neil Russell$^b$\\}
\bigskip
{\hskip 1 truein
$^a$Physics Department\\}
{\hskip 1 truein
Colby College\\}
{\hskip 1 truein
Waterville, ME 04901 USA\\}
\medskip
{\hskip 1 truein
$^b$Physics Department\\}
{\hskip 1 truein
Indiana University\\}
{\hskip 1 truein
Bloomington, IN 47405 USA\\}
\end{flushleft}

\vglue 0.8cm

\noindent
{\bf INTRODUCTION}
\vglue 0.4 cm 

The CPT theorem
\cite{cpt} 
states that
local relativistic quantum field theories of point particles 
in flat spacetime must be invariant under 
the combined operations
of charge conjugation C, parity reversal P, 
and time reversal T.
As a result of this invariance,
particles and antiparticles have equal masses, lifetimes,
charge-to-mass ratios, and gyromagnetic ratios.  
The CPT theorem has been tested to great accuracy in
a variety of experiments
\cite{pdg}.
The best bound is obtained in experiments with neutral mesons,
where the figure of merit is
\beq
r_K \equiv |(m_K - m_{\overline{K}})/m_K| 
\lsim 2 \times 10^{-18}
\quad .
\label{rK}
\eeq
Experiments in Penning traps have also yielded
sharp bounds on CPT violation,
including the best bounds on lepton and baryon systems.
Two types of experimental tests are possible in Penning traps.
Both involve making accurate measurements of cyclotron
frequencies $\om_c$ and anomaly frequencies $\om_a$ of
single isolated particles confined in the trap.
The first compares the ratio $2 \om_a / \om_c$ for particles
and antiparticles.
In the context of conventional QED,
this ratio equals $g-2$ for the particle or antiparticle.
A second experiment compares values of $\om_c \sim q/m$,
where $q>0$ is the magnitude of the charge and $m$ is the mass, 
and is therefore a comparison of charge-to-mass ratios.

Experiments comparing $g-2$ for electrons and positrons
yield the figure of merit
\cite{vd,geo}
\beq
r_g \equiv |(g_{e^-} - g_{e^+})/g_{\rm avg}|
\lsim 2 \times 10^{-12}
\quad ,
\label{rg}
\eeq
while the charge-to-mass-ratio experiments yield the bound
\cite{schwin81}
\beq
r_{q/m}^e \equiv |\left[ (q_{e^-}/m_{e^-})
- (q_{e^+}/m_{e^+})
\right] /(q/m)_{\rm avg}|
\lsim 1.3 \times 10^{-7}
\quad .
\label{rqme}
\eeq

To date, no experiments measuring $g-2$ for protons
or antiprotons have been performed in Penning traps
because of the difficulty in obtaining sufficient
cooling and an adequate signal for detection of the weaker 
magnetic moments.
However,
proposals have been put forward that might make these
types of experiments feasible in the future
\cite{gproton}.
The best current tests of CPT in proton and antiproton systems
come from comparisons of the charge-to-mass ratios
\cite{gg1},
which yield the bound
\beq
r_{q/m}^p \equiv |\left[ (q_p/m_p)
- (q_{\overline{p}}/m_{\overline{p}})
\right]/(q/m)_{\rm avg}|
\lsim
1.5 \times 10^{-9}
\quad .
\label{rqmp}
\eeq

It is interesting to note that in the neutral meson experiments
which yield the bound on $r_K$ in \rf{rK},
measurements are made with an experimental uncertainty
of approximately one part in $10^4$.
In contrast,
measurements of frequencies in Penning traps have
experimental uncertainties of about one part in $10^9$.
This raises some intriguing questions about the Penning-trap
experiments as to why they do not provide better tests
of CPT when they have better experimental precision.
In the context of conventional QED,
which does not permit CPT breaking,
it is not possible to pursue these types of questions.
Instead,
one would need to work in the context of a theoretical
framework that allows CPT breaking,
making possible an investigation of possible experimental
signatures.
Only recently has such a framework been developed
\cite{ck}.

In this paper,
we describe the application of this theoretical framework to
experiments on electron-positron and proton-antiproton
systems in Penning traps.
Our results have been published in 
Refs.\ \cite{bkr1,bkr2}.

\vglue 0.6 cm 
\noindent
{\bf THEORETICAL FRAMEWORK}
\vglue 0.4 cm 

The framework we use 
\cite{ck}
is an extension of 
the SU(3)$\times$SU(2)$\times$U(1) standard model 
originating from the idea of spontaneous CPT and Lorentz breaking
in a more fundamental model such as string theory
\cite{kp,ks}.
This framework preserves various desirable features
of quantum field theory such as
gauge invariance and power-counting renormalizability.
It has two sectors,
one that breaks CPT and one that preserves CPT,
while both break Lorentz symmetry.
The possible CPT and Lorentz violations are parametrized
by quantities that can be bounded by experiments. 
Within this framework,
the modified Dirac equation describing a fermion
with charge $q$ and mass $m$ is given by
\beq
\left( i \ga^\mu D_\mu - m - a_\mu \ga^\mu
- b_\mu \ga_5 \ga^\mu - \half H_{\mu \nu} \si^{\mu \nu} 
+ i c_{\mu \nu} \ga^\mu D^\nu 
+ i d_{\mu \nu} \ga_5 \ga^\mu D^\nu \right) \ps = 0
\quad .
\label{dirac}
\eeq
Here, 
$\ps$ is a four-component spinor,
$i D_\mu \equiv i \partial_\mu - q A_\mu$,
$A^\mu$ is the electromagnetic potential in the trap,
and $a_\mu$, $b_\mu$, 
$H_{\mu \nu}$, $c_{\mu \nu}$, $d_{\mu \nu}$
are the parameters describing possible violations
of CPT and Lorentz symmetry.
The transformation properties of $\ps$ 
imply that the terms involving
$a_\mu$, $b_\mu$ 
break CPT
while those involving 
$H_{\mu \nu}$, $c_{\mu \nu}$, $d_{\mu \nu}$ 
preserve it,
and that Lorentz symmetry is broken by all five terms.
Since no CPT or Lorentz breaking has been observed 
in experiments to date,
the quantities $a_\mu$, $b_\mu$, 
$H_{\mu \nu}$, $c_{\mu \nu}$, $d_{\mu \nu}$ 
must all be small.

\vglue 0.6 cm 
\noindent
{\bf PENNING-TRAP EXPERIMENTS}
\vglue 0.4 cm 

We use this theoretical framework to analyze tests of
CPT and Lorentz symmetry in Penning-trap experiments.
To begin,
we note that the time-derivative couplings in
\rf{dirac} alter the standard procedure for obtaining
a hermitian quantum-mechanical hamiltonian operator.
To overcome this,
we first perform a field redefinition at the lagrangian
level that eliminates the additional time derivatives.
We also use charge conjugation to obtain a Dirac equation
and hamiltonian for the antiparticle.

To test CPT,
experiments compare the cyclotron and anomaly
frequencies of particles and antiparticles.
According to the CPT theorem,
particles and antiparticles of opposite spin in a
Penning trap with the
same magnetic fields but opposite electric fields should
have equal energies.
The experimental relations $g-2 = 2 \om_a / \om_c$ and 
$\om_c = qB/m$ provide connections to the quantities
$g$ and $q/m$ used in defining the figures of merit
$r_g$, $r_{q/m}^e$, and $r_{q/m}^p$.
We perform calculations using 
Eq.\ \rf{dirac} to obtain possible shifts in the
energy levels due to either CPT-breaking or CPT-preserving
Lorentz violation.
In this way,
we examine the effectiveness of Penning-trap
experiments as tests of both CPT-breaking and
CPT-preserving Lorentz violation.
{}From the computed energy shifts we determine how the
frequencies $\om_c$ and $\om_a$ are affected and if
the conventional figures of merit are appropriate.

For experiments in Penning traps,
the dominant contributions to the energy come from
interactions of the particle or antiparticle with the
constant magnetic field of the trap.
The quadrupole electric fields generate smaller effects.
In a perturbative calculation,
the dominant CPT- and Lorentz-violating effects can therefore
be obtained by working with relativistic Landau levels
as unperturbed states.
Conventional perturbations,
such as the anomaly,
will lead to corrections that are the same for particles
and antiparticles.
CPT- and Lorentz-breaking effects will result in
either differences between particles and antiparticles
or in unconventional effects
such as diurnal variations in the measured frequencies.

\vglue 0.6 cm
\noindent
{\bf RESULTS}
\vglue 0.4 cm

Our calculations for electrons and positrons in
Penning traps 
\cite{bkr1}
show that the leading-order effects
due to CPT and Lorentz breaking cause corrections to the
cyclotron and anomaly frequencies:
\beq
\om_c^{e^-} \approx \om_c^{e^+} \approx
(1 - c_{00}^e - c_{11}^e - c_{22}^e) \om_c
\quad ,
\label{wcelec}
\eeq
\beq
\om_a^{e^\mp} \approx \om_a
\mp 2 b_3^e + 2 d_{30}^e m_e + 2 H_{12}^e
\quad .
\label{waelec}
\eeq
Here,
$\om_c$ and $\om_a$ represent the
unperturbed frequencies,
while $\om_c^{e^\mp}$ and $\om_a^{e^\mp}$ denote
the frequencies including the corrections.
Superscripts have also been added on the coefficients
$b_\mu$, etc.\ to denote that these are parameters
of the electron-positron system.
{}From these relations we find the electron-positron differences
for the cyclotron and anomaly frequencies to be
\beq
\De \om_c^e \equiv \om_c^{e^-} - \om_c^{e^+} \approx 0
\quad , \qquad 
\De \om_a^e  \equiv \om_a^{e^-} - \om_a^{e^+} \approx - 4 b_3^e
\quad .
\label{delwce}
\eeq
Evidently,
in the context of this framework comparisons of
cyclotron frequencies to leading order
do not provide a signal for CPT or Lorentz breaking,
since the corrections to $\om_c$ for electrons and
positrons are equal.
On the other hand,
comparisons of $\om_a$ provide unambiguous tests of CPT
since only the CPT-violating term with $b_3$ results in
a nonzero value for the difference $\De \om_a^e$.

We have also found that to leading order there are no
corrections to the $g$ factors for either
electrons or positrons.
This leads to some interesting and unexpected results
concerning the figure of merit $r_g$ in
Eq.\ \rf{rg}.
With $g_{e^-} \approx  g_{e^+}$ to leading order,
we find that $r_g$ vanishes,
which would seem to indicate the absence of CPT violation.
However,
this cannot be true since the model contains explicit CPT violation.
Furthermore,
our calculations show that with $\vec b \ne 0$ the
experimental ratio $2 \om_a / \om_c$ is field dependent and 
is undefined in the limit of vanishing magnetic field.
Thus,
the usual relation $g-2 = 2 \om_a / \om_c$ does not hold
in the presence of CPT breaking.
For these reasons,
the figure of merit $r_g$ in
Eq.\ \rf{rg} is misleading,
and an alternative is suggested.
Since the CPT theorem predicts that states
of opposite spin in the same magnetic field have equal energies,
we propose as a model-independent figure of merit,
\beq
r^e_{\om_a}
\equiv \fr{|{E}_{n,s}^{e^-} - {E}_{n,-s}^{e^+}|}
{{E}_{n,s}^{e^-} }
\quad ,
\label{re}
\eeq
where ${E}_{n,s}^{e^\mp}$ are the energies of the
relativistic states labeled by their Landau-level 
numbers $n$ and spin $s$.
Our calculations show $r^e_{\om_a} \approx 
| \De \om_a^e | / 2 m_e  \approx |2 b_3^e | / m_e$,
and we estimate as a bound on this figure of merit,
\beq
r^e_{\om_a} \lsim 10^{-20}
\quad .
\label{relim}
\eeq
In Ref.\ \cite{bkr2},
we describe additional possible signatures 
of CPT and Lorentz breaking.
These include possible diurnal variations in the anomaly
and cyclotron frequencies.
Tests for these effects would provide bounds 
on various components of the
parameters $c_{\mu \nu}^e$, $d_{\mu \nu}^e$, and $H_{\mu \nu}^e$
at a level of about one part in $10^{18}$.

A similar analysis can also be performed on proton-antiproton
experiments in Penning traps.
In this context,
it suffices to work at the level of an effective theory
in which the protons and antiprotons are regarded as
basic objects described by a Dirac equation.
The coefficients $a_\mu^p$, $b_\mu^p$, $H_{\mu \nu}^p$, 
$c_{\mu \nu}^p$, $d_{\mu \nu}^p$ represent effective
parameters,
which at a more fundamental level depend on the underlying
quark interactions.
Comparisons of protons and antiprotons in the context
of this model yield the results for the proton-antiproton
frequency differences,
\beq
\De \om_c^p \equiv \om_c^{p} - \om_c^{\bar p} = 0
\quad , \qquad
\De \om_a^p  \equiv \om_a^{p} - \om_a^{\bar p} = 4 b_3^p
\quad .
\label{delwap}
\eeq
Assuming an experiment could be made sensitive enough
to measure $\om_a^p$ and $\om_a^{\bar p}$ with a precision
similar to that of electron $g-2$ experiments,
then the appropriate figure of merit would be
\beq
r^p_{\om_a}
\equiv \fr{|{ E}_{n,s}^{p} - { E}_{n,-s}^{\bar p}|} 
{{ E}_{n,s}^{p}}
\quad .
\label{rp}
\eeq
A bound on this can be estimated as
\beq
r^p_{\om_a}
\lsim 10^{-23}
\quad .
\label{rp3}
\eeq
It is apparent that an experiment comparing anomaly
frequencies of protons and antiprotons in a Penning trap
has the potential to provide a particularly tight CPT bound.
Other signatures of CPT and Lorentz breaking involving
diurnal variations in $\om_a$ and $\om_c$ are described in
Ref.\ \cite{bkr2}.
These additional signatures provide bounds on various components of
$c_{\mu \nu}^p$, $d_{\mu \nu}^p$, and $H_{\mu \nu}^p$
estimated at about one part in $10^{21}$.

\vglue 0.6 cm
\noindent
{\bf CONCLUSIONS}
\vglue 0.4 cm

We find that the use of a general theoretical framework
incorporating CPT and Lorentz breaking permits a detailed
investigation of possible experimental signatures in 
Penning-trap experiments.
Our results indicate that the sharpest tests of CPT symmetry
emerge from comparisons of anomaly frequencies in $g-2$ experiments.
Our estimates of appropriate figures of merit provide bounds of
approximately $10^{-20}$ in electron-positron experiments
and of $10^{-23}$ for a plausible proton-antiproton experiment.
Other signals involving possible diurnal variations provide
additional bounds at the level of $10^{-18}$ in the
electron-positron system 
and $10^{-21}$ in the proton-antiproton system.
A table showing all our estimated bounds is presented in 
Ref.\ \cite{bkr2}.

\vglue 0.6 cm
\noindent
{\bf ACKNOWLEDGMENTS}
\vglue 0.4 cm

This work was supported in part
by the National Science Foundation 
under grant number PHY-9503756.

\vglue 0.6 cm
\noindent
{\bf REFERENCES}
\vglue 0.4 cm

\end{document}